\documentclass[12pt,a4paper]{article}
\usepackage{graphicx, wrapfig}
\usepackage{times}
\usepackage{xcolor}

\def\wasp{{\it SuperWASP}}
\def\hipp{{\it Hipparcos}}

\def\rsun{$\rm{R}_{\odot}$}           

\def\deg{\hbox{$^\circ$}}              

\newcommand{\ADDFIG}[1]{} 
\newcommand{\ADDTAB}[1]{} 
\newcommand{\ADDWORK}[1]{} 
\title{\bf Magnetic activities on active solar-type stars}
\author{Subhajeet Karmakar$^{1,2}$\thanks{subhajeet@prl.res.in, subhajeet09@gmail.com}, Jeewan C. Pandey$^{2}$, Sachindra Naik$^{1}$, \\ Igor S. Savanov$^{3}$, and Ashish Raj$^{4}$ \\
\vspace{0.5cm}\\
\normalsize $^1$ Physical Research Laboratory, Navrangapura, Ahmedabad 380009, India\\ 
\normalsize $^2$ Aryabhatta Research Institute of observational Sciences , Nainital 263002, India \\
\normalsize $^3$ Institute of Astronomy, Russian Academy of Sciences, ul. Pyatniskaya 48, Moscow 119017, Russia\\
\normalsize $^4$ Indian Institute of Astrophysics, II Block Koramangala, Bangalore 560034, India
}
\date{\mbox{}}
\begin{document}
\maketitle
\setcounter{page}{1001}
\pagestyle{plain}
    \makeatletter
    \renewcommand*{\pagenumbering}[1]{%
       \gdef\thepage{\csname @#1\endcsname\c@page}%
    }
    \makeatother
\pagenumbering{arabic}

%
%
\def\bull{\vrule height .9ex width .8ex depth -.1ex}
\makeatletter
\def\ps@plain{\let\@mkboth\gobbletwo
\def\@oddhead{}\def\@oddfoot{\hfil\scriptsize\bull\quad
"2nd Belgo-Indian Network for Astronomy \& astrophysics (BINA) workshop'', held in Brussels (Belgium), 9-12 October 2018 \quad\bull}%
\def\@evenhead{}\let\@evenfoot\@oddfoot}
\makeatother
%
%
\def\beginrefer{\section*{References}%
\begin{quotation}\mbox{}\par}
\def\refer#1\par{{\setlength{\parindent}{-\leftmargin}\indent#1\par}}
\def\endrefer{\end{quotation}}
%
%
{\noindent\small{\bf Abstract:}
We present results obtained from the studies of magnetic activities on four 
solar-type stars (F-type star KIC~6791060, K-type star LO Peg, and two 
M-type planet-hosting stars K2--33 and EPIC 211901114) by using optical 
observations from several ground- and space-based telescopes. In this study, 
we investigate magnetic activities such as spot-topographic evolution 
and flaring events in these stars. We compare the results 
obtained from this study with that of the Sun. In the surface temperature maps, 
one active longitude has been detected in KIC~6791060, whereas in each of the 
other three cases two active longitudes are seen. The spottedness was found 
to vary in the range of 0.07--0.44\%, 9--26\%, 3.6--4.2\%, and 4.5--5.3\% for KIC~6791060, 
LO Peg, K2--33, and EPIC 211901114, respectively. Several flaring events have 
been identified in each star. An increasing trend in flaring frequency per stellar 
rotation has been found in the stars with change in spectral type from F to M. 
These findings indicate the increase in magnetic activities with the spectral type
of stars. This can be explained due to increasing the ratio of the thickness of the 
convection zone to the radiation zone from F-type star to the M-type stars.
}
\vspace{0.5cm}
\section{Introduction}

\label{sec:intro}

Cool stars with spectral type from late-F to early-M have a similar internal structure 
to that of the Sun and are called as solar-type stars. In these stars, Hydrogen burning at the core is the process of energy generation. The energy is being transported outward
through radiation (inner part of the star) and convection (outer part of the star) depending on the steepness of the temperature gradient. The interface between the radiation and the convection 
zones is called the ``tachocline''. According to the current understanding of stellar dynamo 
theory, the magnetic field in the Sun is supposed to be generated 
by a naturally occurring dynamo mechanism operating inside. 
Solar-type stars with a similar internal structure to that of the Sun are 
expected to operate a similar kind of dynamo mechanism. However, the observational
evidence of range of stellar rotation periods, surface gravities, masses and ages 
which put into the debate on the existing magnetic dynamo theory (see Favata et al. 2000, Karmakar et al. 2017). 
The observational evidence of the dynamo generated activities are presence of dark spots, flaring events, etc. 
In order to provide useful constraints for the dynamo theory, we are currently working 
on a project to investigate the magnetic activities in the stars with the same internal structure. 
We have selected an F-type star KIC 6791060, a K-type star LO Peg, and two M-type 
planet-hosting stars K2--33 and EPIC 211901114 to investigate the magnetic activities
and compare the results with that of the Sun. 

The basic parameters of the solar-type stars under studied are given in 
Table~\ref{tab:parameter}. KIC~6791060 is a poorly-known F5-type, main-sequence 
(Frasca et al. 2016), ultra-fast rotator (UFR) with a rotational period of 0.344 day. 
Balona (2015) has categorized this star as a rotationally variable star with $V$ 
magnitude of 10.57. Using the Gaia Data Release~2 (DR2), Gaia Collaboration (2018) derived 
parallax of this star to be 3.41$\pm$0.02 mas which 
corresponds to a distance  of 290.7 pc by Bailer-Jones et al. (2018). LO~Peg
is an active, single, young, main-sequence, K5--8 type UFR with a rotational 
period of 0.4231$\pm$0.0001 day (Karmakar et al. 2016). LO~Peg has been extensively 
studied in last two decades. Strong flaring activities in this star are identified 
from H$\alpha$ and He~I D3 observations (Jeffries et al. 1994, Eibe et al. 1999) 
as well as from optical observations (Ta\c{s} 2011). Doppler images of LO~Peg 
provided evidence of high polar activities (Piluso et al. 2008 and references 
therein). Several photometric, polarimetric, and X-ray studies also carried out 
by Dal \& Ta\c{s} (2003), Pandey et al. (2005), Csorv{\'a}si (2006), Ta\c{s} (2011), 
Karmakar et al. (2016), and Savanov et al. (2016).

%
\renewcommand{\tabcolsep}{8mm}
\begin{table}
\caption{Properties of selected stars}
\label{tab:parameter}
\begin{tabular}{llccl}
\hline
Object             & Spectral Type             & V            &Period            &Distance \\
                 &                           & (mag)        &(day)            &(pc)\\
\hline
KIC~6791060      &    F5                     &  10.57      & 0.344            & 290.7 \\
LO~Peg           &    K5--8                  &   9.25      & 0.4231           &  25.1 \\
K2--33            &    M2                     &  15.71      & 6.2895           & 139.3 \\
EPIC~211901114   &    M2.7                   &  16.85       & 8.6075           & 184.0 \\
\hline
\end{tabular}
\vspace{-0.5cm}
\end{table}
 
 K2--33 is an extremely young, pre-main-sequence star of spectral type M2 
 located at a distance of  139.3 pc in the Upper Scorpius subgroup
 of the Scorpius--Centaurus Association (Mann et al. 2016). 
 The age of this subgroup has been estimated to be 
11$\pm$2 million years (Pecaut et al. 2012), while evolutionary models estimated 
the age of K2--33 as 9.3 million years (David et al. 2016). K2--33 is known to host
one Neptune-sized exoplanet. The parallax measured for this star by 
Gaia DR2 is 7.15$\pm$0.08 mas (Gaia Collaboration 2018). The estimated 
radius for K2--33 is 1.05$\pm$0.07 \rsun\ (Mann et al. 2016).
EPIC~211901114 is an active red-dwarf of spectral type M2.7 
(Kraus et al. 2007) hosts a Jupiter size exoplanet (Rebull et al. 2017). 
EPIC~211901114 is a part of the Praesepe cluster and having an age of 
$\sim$790 million years (Rebull et al. 2017).
Gaia Collaboration (2018) estimated the parallax of EPIC~211901114 
to be 5.41$\pm$0.08 mas. Using Gaia DR2 Bailer-Jones et al. (2018)
estimated a distance of EPIC~211901114 to be  184 pc.

The paper is structured  as follows: in Section~2, we describe the observations that 
have been used in our analysis and the methods of data reduction. Section~3 describes 
data analysis and results obtained from our work and in Section~4, we present our 
discussion and conclusions.

\vspace{-0.3mm}
\section{Observations and Data Reduction}
\label{sec:obs}

Observations of the selected stars have been carried out by using several 
ground and space-based observatories. KIC~6791060 was in the field that 
was observed by the first Kepler mission for four years starting from 2009, 
December 13 to 2013, December 13. The Kepler data were obtained in two 
different modes. In eighteen quarters, the observations were taken in 
long-cadence (LC mode; 30-min exposures) and in one quarter, the field 
was observed only in short-cadence (SC mode; 1-min exposures).
We observed LO~Peg in $U$, $B$, $V$, and $R$ photometric bands
using  6-m Special Astrophysical Observatory of RAS,
2-m IUCAA Girawali Observatory Telescope, 1.04-m 
ARIES Sampurnanand Telescope, 0.5-m Zvenigorod Observatory of INASAN, 
and 0.36-m Goddard Robotic Telescope. The exposure time was between 5 and 60 s depending on 
the seeing condition, filter, and telescope used. Apart from this, we have also used data
from literature and archive. The details of the sources of $V$ band data are given in 
Karmakar et al. (2016) and Savanov et al. (2016).
\begin{figure}[t]
\centering
\includegraphics[width=14.1cm,]{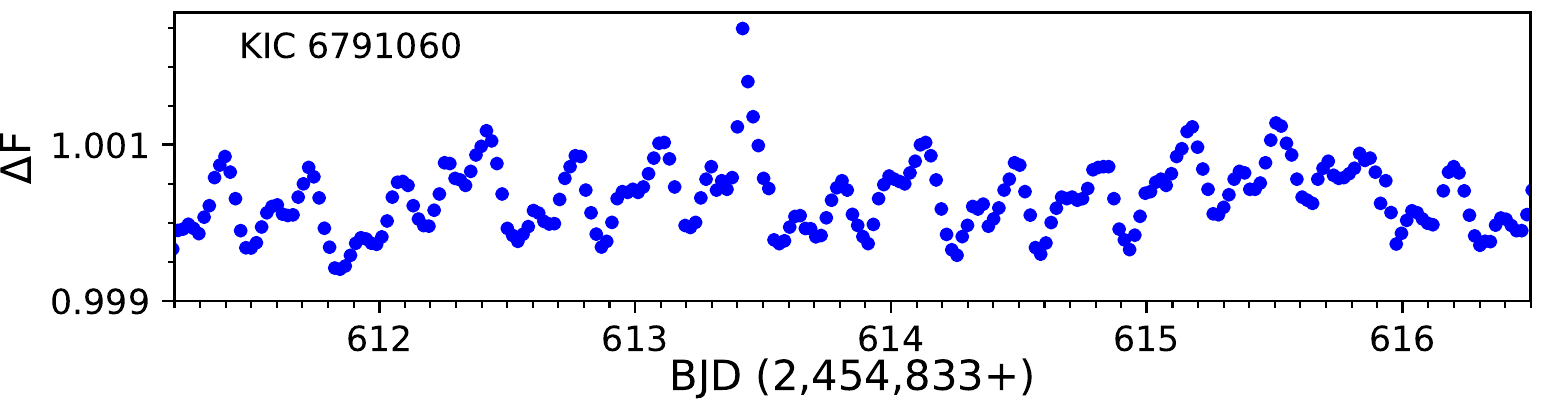}
\includegraphics[width=14.1cm,]{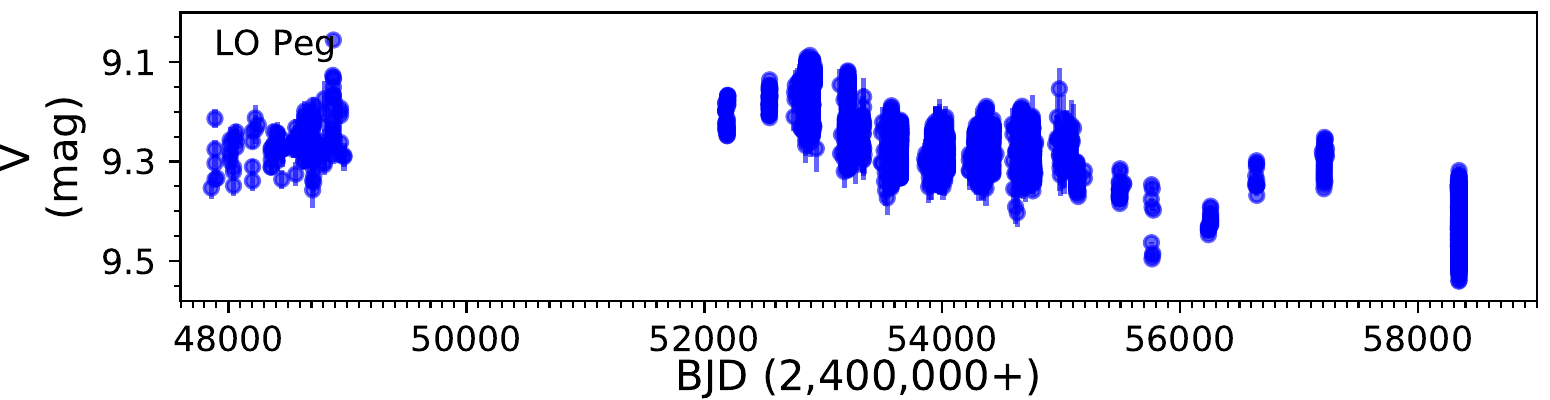}
\includegraphics[width=14.1cm,]{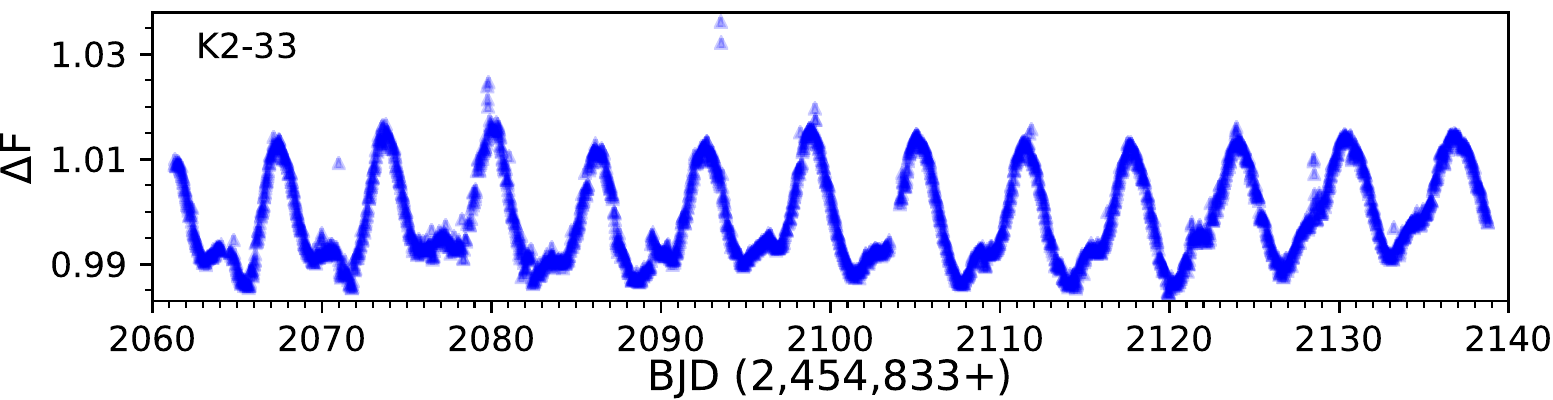}
\includegraphics[width=14.1cm,]{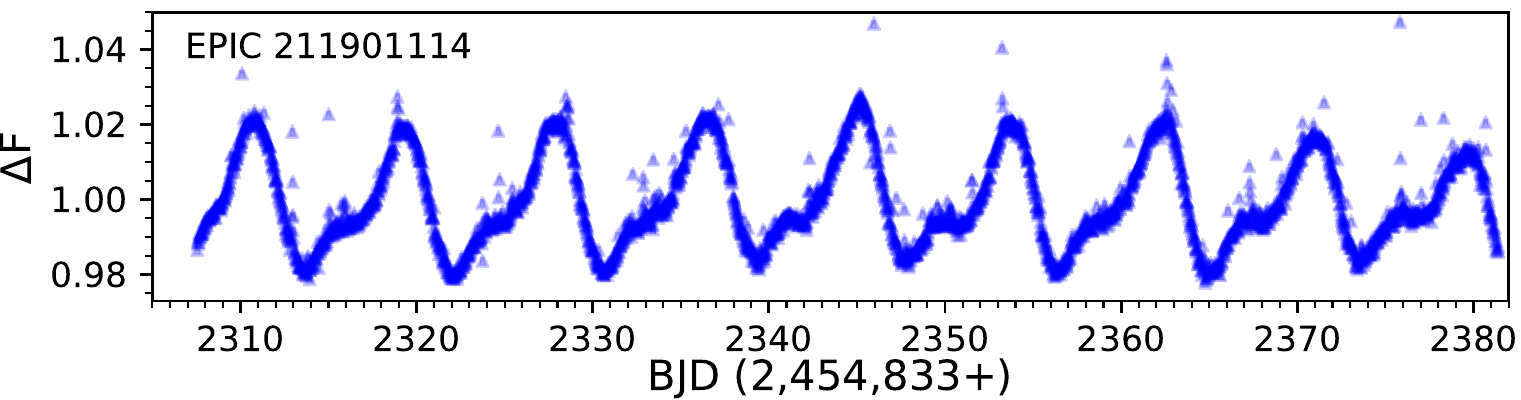}
\caption{Optical light curves of selected four objects. 
Normalized source flux of KIC~6791060 (top panel) was obtained
from first Kepler mission, second panel shows the  $V$ band 
light curve of LO~Peg. The details of observations
of $V$ band data are given in Karmakar et al. (2016).
The normalized flux for K2--33 (third panel) and EPIC~211901114 (fourth panel)
were obtained from K2 mission.
\label{fig:lc}}
\end{figure}

\vspace{-1.3mm}
\section{Analysis and results}

\subsection{Optical light curves}

The optical light curves of the four objects are shown in Fig.~\ref{fig:lc}. 
The top panel shows a part of the light curve of KIC~6791060 in the unit of normalized 
flux as observed with Kepler satellite. The second panel of Fig.~\ref{fig:lc} 
shows the $\sim$29 years long $V$-band light curve of the K-type star LO~Peg. 
The third and fourth panels show seventy days long light curves of K2--33 and EPIC 211901114, 
respectively. 
The light curves of KIC~6791060 and LO Peg display large variation in amplitudes both
in short- and long-term time scales, whereas  due to relatively smaller duration of
observation, long-term variation in K2--33 and EPIC~211901114 is not well evident.
However, in both cases, short time-scale variations have been seen.

We search for periodicities in all four objects using Scargle--Press period 
search method (Scargle 1982; Horne \& Baliunas 1986; Press \& Rybicki 1989).
For KIC~6791060, in the Scargle--Press power spectrum the highest peak 
corresponds to a period of 0.34365$\pm$0.00004 day. 
This period could be the rotation period of the star. This derived period for KIC~6791060  
agrees with previous determinations (Balona 2015, McQuillan et al. 2014).
 In the power spectrum, another peak was also seen which corresponds to a period
 of 0.34349$\pm$0.00004 day, and is separated by 0.00016 day from the rotational period.
 This multiple periodicity in the light curve results in a beat pattern in the four year 
 long light curve near $\sim$550 and $\sim$1350 day in time = BJD -- 2\,454\,833 day; 
 (see bottom panel of Fig.~1 of Karmakar et al. 2018).
This multiple periodicity could be a result of multiple spot groups located
in different latitude.
From the $V$-band data of LO~Peg we estimated the period to be 0.4231$\pm$0.0001 day.
 In our analysis we have also calculated  the False Alarm Probability  
 for any peak frequency using the method given by Horne \& Baliunas (1986).
 Further, we have phase folded the data in each  period that was detected above 99.9\%
 significance level and found periodic modulation only for periods 5.98 and 2.2 year. 
 The strongest peak in the power spectra of K2--33 and EPIC~211901114 
 corresponding to a period of 6.29$\pm$0.50 and  8.56$\pm$0.60 day, respectively.
Mann et al. (2017) give the results of an independent determination of 
the stellar rotation periods, which agree with our derived values.
Several flaring events were also detected during the observations of all
four above mentioned stellar systems which will be discussed 
in Section~\ref{sec:flares}.

\subsection{Stellar surface imaging}
\label{subsec:si_li}

Starspots move across the stellar disk with the rotation and thus modulate 
the total brightness. The inversion of the phased light curves into stellar 
images allow us to determine the locations of spots on the stellar surface. 
In order to do light curve inversion, we have used i\textsc{ph} code (Savanov 
\& Strassmeier 2008). According to the basic assumption of the model, the 
local intensity ($I$) of the stellar surface is the sum of the intensity of the 
photosphere ($I_P$) and from the cool spots ($I_S$) weighted by the spot
filling factor $f$ through the following relation $I = f \times I_P + (1 - f ) 
\times I_S$ ; with $0 < f < 1$. Here, the spot filling factor $f$ is 
defined as the fraction of the stellar surface covered by spots.

\begin{figure}[t]
\rotatebox[origin=c]{90}{\textcolor{white}{\ldots}\Large Latitude~(\deg)}
\vspace{-0.5mm}
\begin{minipage}{8.0cm}
\centering
\includegraphics[width=8.0cm, height=4.0cm]{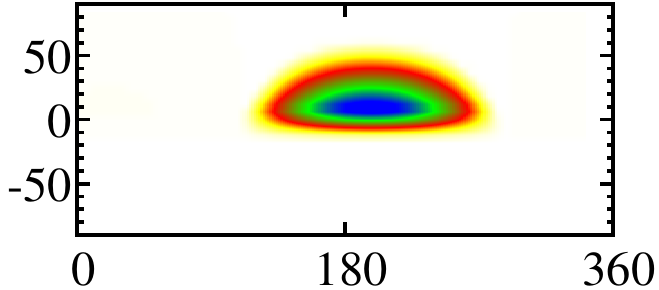}
\end{minipage}
\rotatebox[origin=c]{90}{\textcolor{white}{\ldots}\Large Latitude~(\deg)}
\begin{minipage}{8.0cm}
\centering
\includegraphics[width=8.0cm, height=4.25cm]{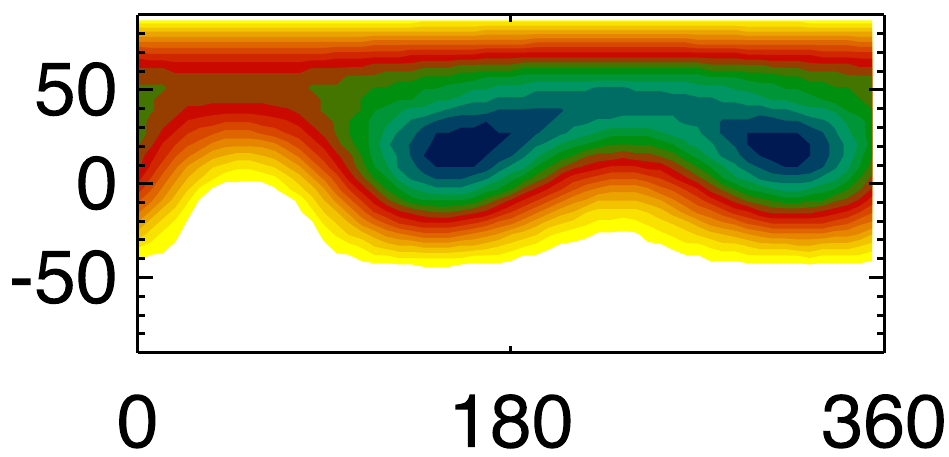}
\end{minipage}
{\Large \textcolor{white}{.}~~~~~~~~~~~~~~~~~~~~~~~Longitude~(\deg)~~~~~~~~~~~~~~~~~~~~~~~~~~~~~~~~~~~~Longitude~(\deg)}\\
\rotatebox[origin=c]{90}{\textcolor{white}{\ldots}\Large Latitude~(\deg)}
\vspace{-0.8mm}
\begin{minipage}{8.0cm}
\centering
\includegraphics[width=8.0cm, height=4.0cm]{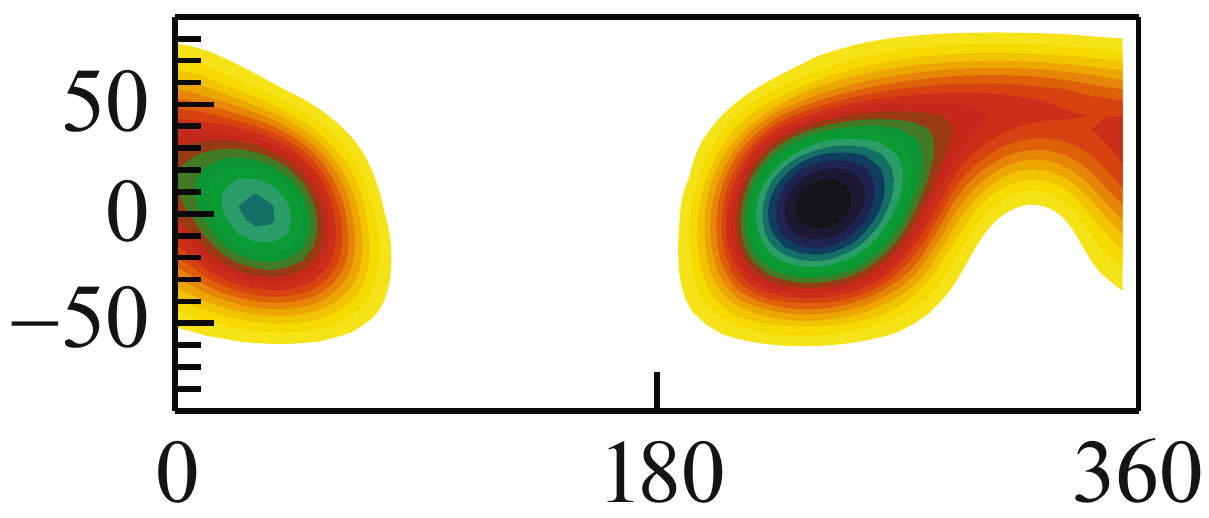}
\end{minipage}
\rotatebox[origin=c]{90}{\textcolor{white}{\ldots}\Large Latitude~(\deg)}
\begin{minipage}{8.0cm}
\centering
\includegraphics[width=8.0cm, height=4.0cm]{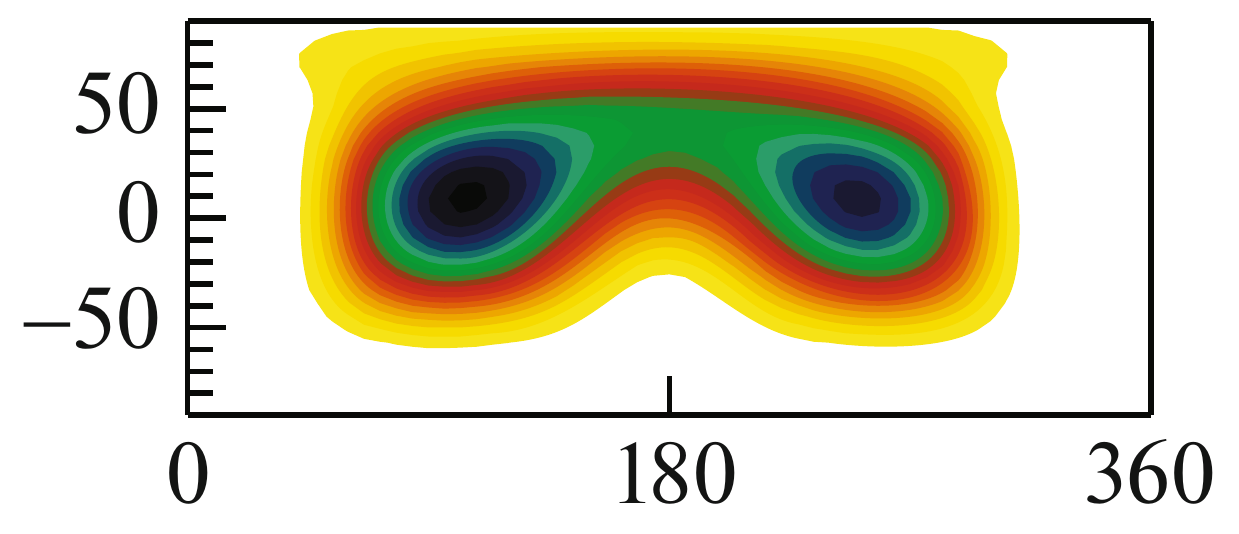}
\end{minipage}
{\Large \textcolor{white}{.}~~~~~~~~~~~~~~~~~~~~~~~Longitude~(\deg)~~~~~~~~~~~~~~~~~~~~~~~~~~~~~~~~~~~~Longitude~(\deg)}\\
\vspace{-5mm}
\caption{Representative surface temperature maps for KIC~6791060 (top left), 
LO~Peg (top right), K2--33 (bottom left), and EPIC~211901114 (bottom right). 
 \label{fig:surface_li}}
\vspace{-3mm}
\end{figure}

In the case of KIC~6791060, considering the duration of observation and 
its rotation period we made 3899 time-intervals (each interval corresponds to a 
single rotation of the star) to construct surface images. For LO~Peg, we made 47 
time intervals by manual inspection in such a way that each interval had sufficient 
number of data points without any noticeable changes in the shape of light curve.
For K2--33 and EPIC~211901114, we divided all the observational data into 12 and 
8 data sets, respectively, each successively covering one rotation period of the star.
In each case, the individual light curves were analyzed using the i\textsc{ph} 
code. In our modeling, the surface of the star was divided into a grid of 
$6^{\circ}\times 6^{\circ}$ pixels and the values of $f$ were determined for 
each grid pixel. We also considered the respective photospheric temperature 
and inclination. We adopted surface temperatures of
KIC~6791060, LO~Peg, K2--33, and EPIC~211901114 to be 6343, 4500, 3540, and 3440~K 
(Luo et al. 2016, Pandey et al. 2005, Mann et al. 2016, 2017) 
for our study. Spot temperature were assumed to be 1000 K lower than photospheric
temperature (Savanov \& Strassmeier 2008). The stellar astrophysical input 
parameters include a set of photometric fluxes calculated from 
an atmospheric model by Kurucz (1992) as a function of temperature and gravity.
Representative surface maps for all four objects are shown 
in Fig.~\ref{fig:surface_li}. 

In case of KIC~6791060, we have found only one active longitude region. Over time, 
we found a shift in the position of the active region along the longitude during 
the four years observation of KIC~6791060. However, in the case of LO~Peg, we found that
the surface temperature maps contain clusters of spots at two longitudes. During 
the $\sim$29 years long observation of LO~Peg, we have also detected the flipping 
of an active longitude region to a less active region by less than 180$\deg$ along 
the longitude also known as flip-flop phenomenon. In case of K2--33 and EPIC~211901114,
all the surface temperature maps contain clusters of spots at two longitudes 
which we represented as two active regions. For EPIC~211901114, we 
saw no relative variations of the position of the active region on
the surface of the star. However, we did find such variations for K2--33. 
The estimated values of spottedness i.e. the total area of the visible surface 
covered by the spots for  KIC~6791060, LO~Peg, K2--33, and EPIC~211901114 vary in 
the range of  0.07--0.44\%, 9--26\%,  3.6--4.2\%,  4.5--5.3\%, respectively and is
further discussed in Section~\ref{sec:discussion}.

\vspace{-0.3mm}
\subsection{Flaring events}
\label{sec:flares}

Flaring events are the explosions on the stellar surface which releases huge 
amount of magnetic energy stored near the starspots into the outer atmosphere. 
In the light curves, the flaring events are identified as positive flux excursions
throughout the data (see Fig.~\ref{fig:lc}). In our flare analysis, we took great 
care to identify it correctly so that we do not miss any small flare and at the 
same time, do not consider an outlier as a flare. Due to the presence of
homogeneous temporal resolution throughout the observation 
of KIC~6791060, K2--33, and EPIC~211901114, we have used the following approach.
In the beginning, the light curves were detrended by fitting the sinusoids with 
periods as obtained from the period analysis. Then, we computed the moving 
standard deviation ($\sigma$) of the detrended light curve. The data-length were taken as 
twice of the rotational period of the star. A large deviation in the data points 
from the detrended light curve gives a larger value of $\sigma$ and were 
flagged as flare candidates. Finally, we manually verified all the flagged flare 
candidates and identified as flares only when 3 consecutive points lie above 
the 2.5$\sigma$ level. 

\begin{figure}[h]
\begin{minipage}{8.2cm}
\centering
\includegraphics[width=8.2cm]{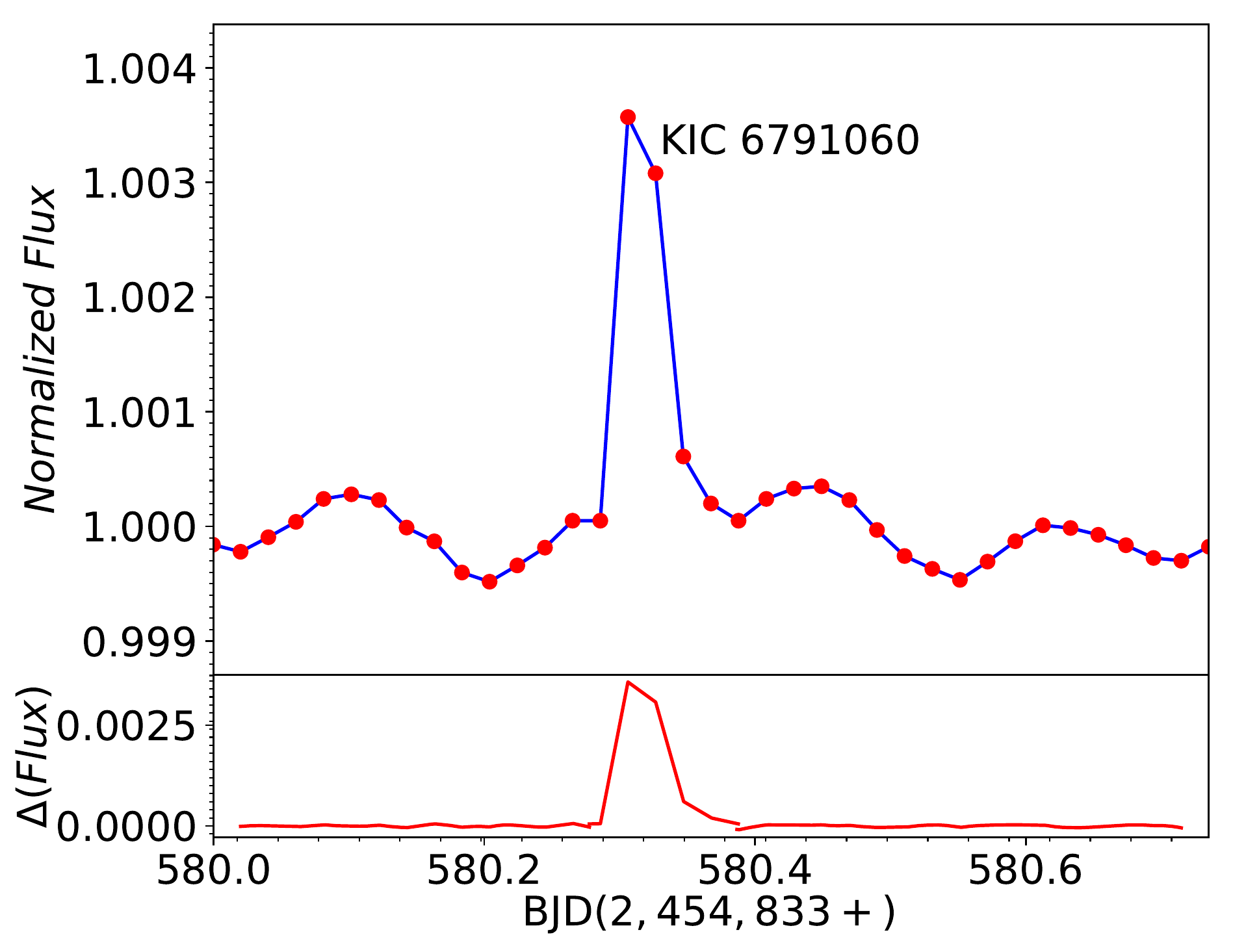}
\end{minipage}
\hfill
\begin{minipage}{8.2cm}
\centering
\includegraphics[width=8.2cm]{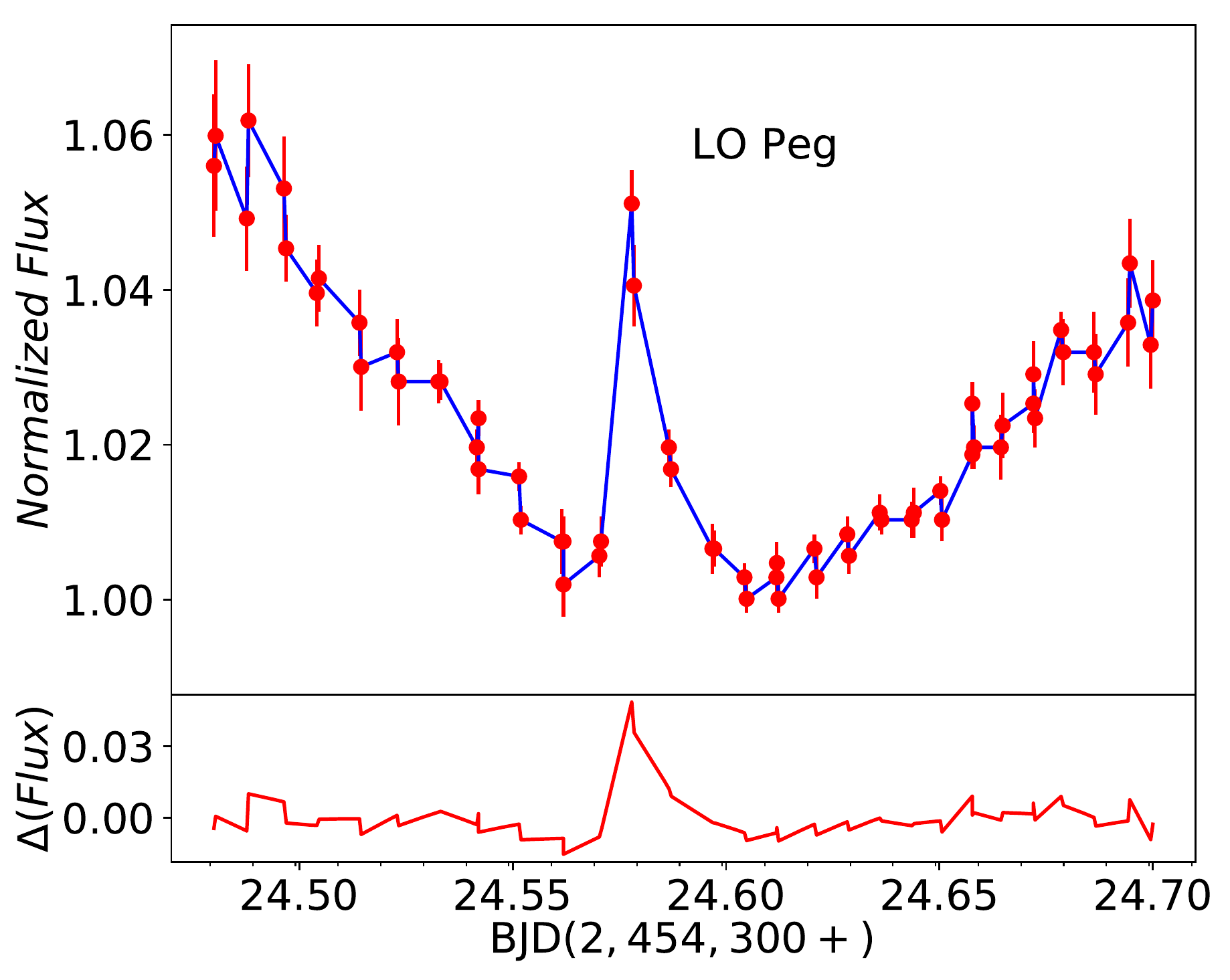}
\end{minipage}
\begin{minipage}{8.2cm}
\centering
\includegraphics[width=8.2cm]{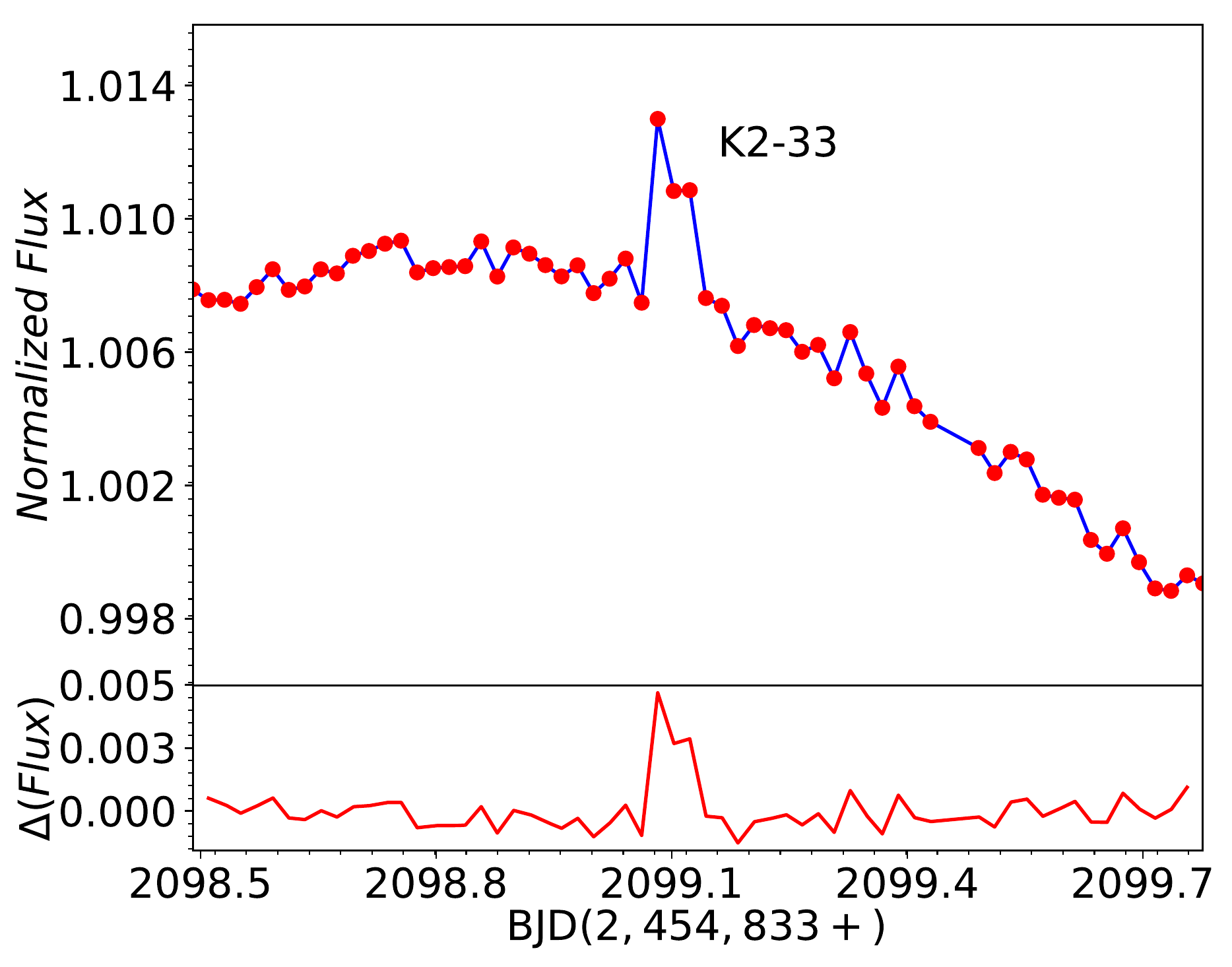}
\end{minipage}
\hfill
\begin{minipage}{8.2cm}
\centering
\includegraphics[width=8.2cm]{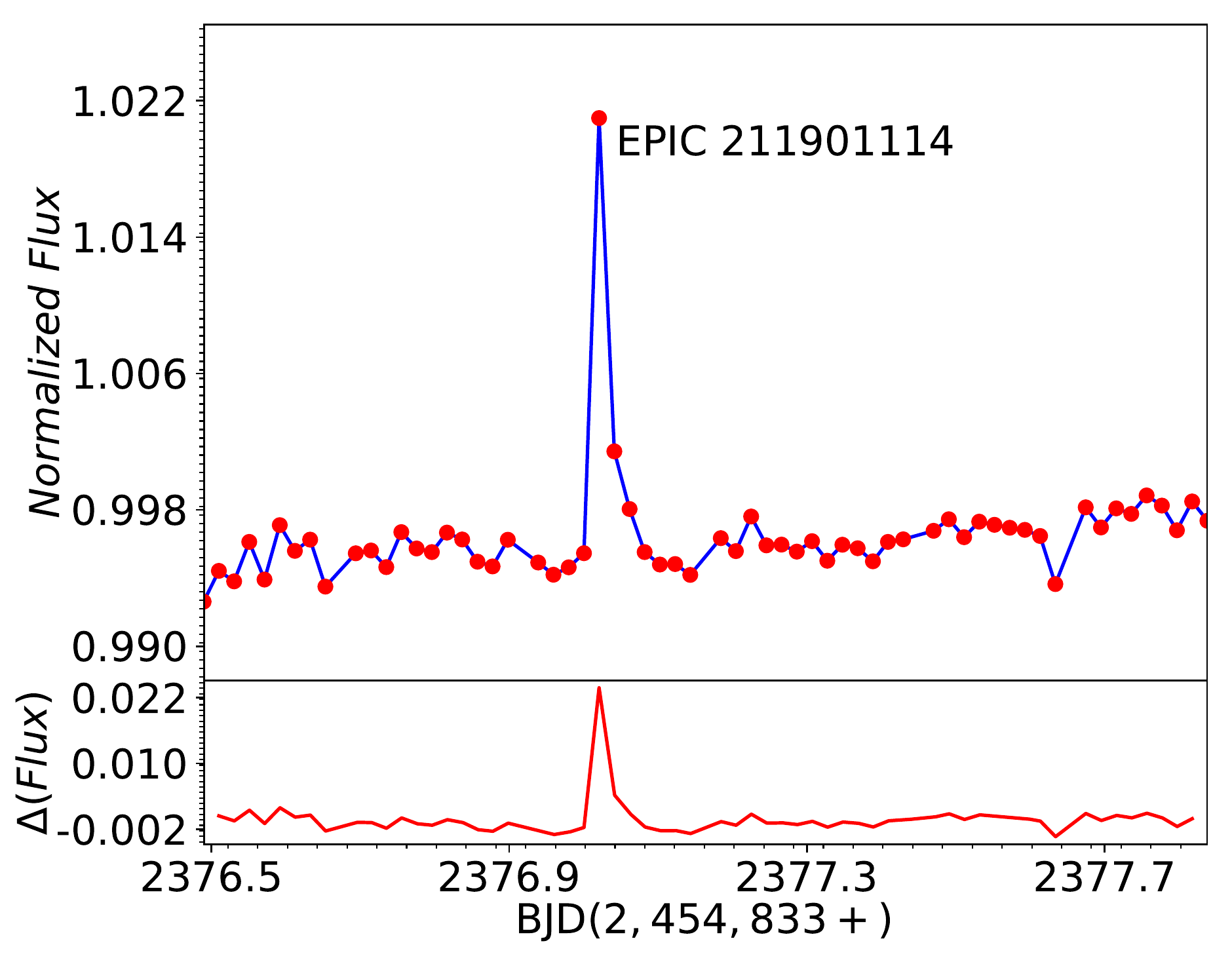}
\end{minipage}
\caption{Light curves of representative flaring events of KIC~6791060 (top left), LO~Peg (top right), K2--33 (bottom-left) and EPIC~211901114 (bottom right). Top panel of each plot shows the light curve which contain a flare. The bottom panel shows the corresponding detrended light curve.
\label{fig:flares}}
\end{figure}

In case of LO~Peg, as the light curve data were compiled with various ground and 
space-based observatory with different cadence and different uncertainties, we followed a
different approach. We chose different epochs in such a way that each epoch contains a continuous single night observation with at least 16 data points and minimum observing 
span of $\sim$1 hour.  A total of 501 epochs were found among which only 82 epochs have simultaneous observations with other three optical bands. The flare identification is carried out following the above criteria but with visual inspection of each epoch separately. Fig.~\ref{fig:flares} shows the light curve of representative flares of 
KIC~6791060, LO~Peg, K2--33, and EPIC~211901114. The top panel of 
each figure shows the light variation with time, whereas the bottom panel shows the detrended light curve. We have detected 38, 20, 7, and 32 flares for KIC~6791060, LO~Peg, K2--33, and EPIC~211901114. 

The flare energy is computed using the area under the flare in light curve
i.e. the integrated excess flux ($F_{e}(t)$) released during the flare (from 
flare start-time $t_s$ to flare end-time $t_e$) as
\begin{equation} 
E_{flare} = 4\pi d^{2} \int_{t_s}^{t_e} \! F_{e}(t) \, dt
\label{eqn:energy}
\end{equation}

\noindent Using the respective distances (d) mentioned in Table~\ref{tab:parameter}, we derived 
the flare energy ($E_{flare}$) for KIC~6791060 in the order of 10$^{31-33}$ ergs. For LO~Peg, the 
flare energies are found to be in the range of 10$^{30.9-34.2}$ ergs.
There are very few detailed studies of optical flares on UFRs due
to constraints in their detection limit, detection 
timing, and observational bias. Our estimates of energy yielded 10$^{32.2-33.3}$ ergs 
for K2--33 and 10$^{32.1-33.4}$ ergs for EPIC~211901114.


\vspace{-1.3mm}
\section{Discussion and conclusions}
\label{sec:discussion}
In this paper, we have presented the magnetic activities of four active solar-type 
stars with spectral type F, K, and M. We have constructed maps of temperature 
inhomogeneities on the surfaces of KIC~6791060, LO~Peg, K2--33, and EPIC~211901114.
As mentioned in Section~\ref{subsec:si_li}, we have detected the flip-flop phenomena
on LO~Peg. The flip-flop has also been noticed by several authors
such as Jetsu et al. (1991), Korhonen et al. (2002).
This phenomenon is well explained by the dynamo based solution 
where a non-axisymmetric dynamo component with two 
permanent active longitudes (180$\deg$ apart) is needed along with an oscillating 
axisymmetric magnetic field. Fluri \& Berdyugina (2004) suggest another possibility 
with a combination of stationary axisymmetric and varying non-axisymmetric components.

\begin{wrapfigure}{r}{0.6\textwidth} 
\vspace{-20pt}
  \begin{center}
    \includegraphics[width=0.6\textwidth]{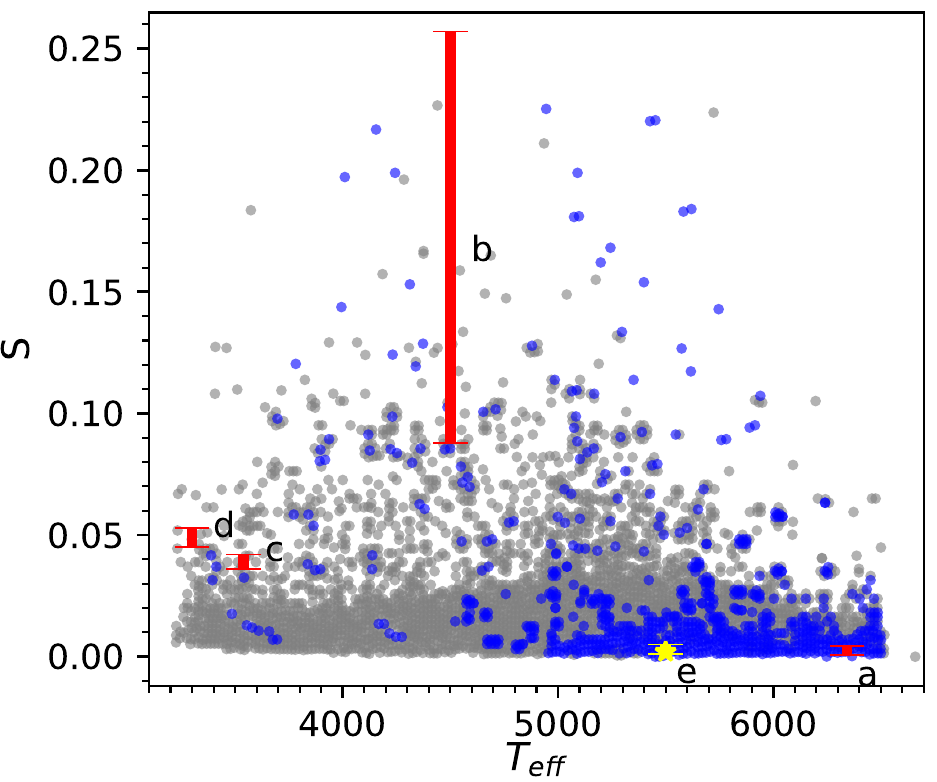}
    \caption{The spottedness (S) vs. stellar effective temperature ($T_{eff}$) plot  for F--M type stars taken from the catalog of McQuillan et al. (2014). The blue and gray solid circles represent the fast- and slow-rotating stars with period $<$1 day and $\geq$1 day, respectively. The red lines marked with (a), (b), (c), and (d) represent the spottedness range discussed in this paper for the stars KIC~6791060, LO~Peg, K2--33, and EPIC~211901114, respectively. The yellow solid star symbol (e) represents the range of solar spottedness.}
    \label{fig:sp_vs_teff}
  \end{center}
  \vspace{-20pt}
  \vspace{1pt}
\end{wrapfigure}
From the starspot imaging, we have also derived  the spottedness of these four stars.
In Fig~\ref{fig:sp_vs_teff} we have shown the spottedness of all
the F--M type stars from McQuillan et al. (2014) along with the Sun and the
stars were investigated in this paper.  This shows that the F-type UFR shows
low level of activity in terms of spottedness than K-type UFR LO~Peg.
However, LO~Peg shows more activity than other two M-type stars
in our sample that could be due to 15--20 times faster rotation rate of 
LO~Peg than those M-type stars. Similar level of magnetic activity 
was also observed in other K-M type stars such as AB~Dor 
(Collier Cameron \& Donati 2002), BE~Cet, DX~Leo, and LQ~Hya (Messina \& Guinan 2003).

A total of 38 flares were detected in KIC~6791060 with a flare frequency of 1 flare per 
40 days (or 1 flare per 116 rotations). In case of LO~Peg, we have 
detected a total of 20 optical flares with a rate of $\sim$1 flare per two 
days (or 1 flare per 5 rotations). 
From K2--33 and EPIC~211901114, the number of total detected flares were 7 and 32, respectively. 
The flare frequency for K2--33 was derived to be $\sim$1 flare per 8 days
(or $\sim$1.2 flares per rotation), whereas in case of EPIC~211901114, 
it was estimated to be $\sim$1 flare per 2 days (or 4.6 flares per 
rotation). Thus, we found an increasing rate in the flare occurrence per rotation
with changing spectral type from F-type to M-type stars.
In case of the Sun, the frequency of occurrence of solar flares varies from 
several flares per day when the Sun is particularly ``active'' to less than one flare 
every week when the Sun is ``quiet'', following the 11-year solar cycle (Benz \& G\"{u}del 2010, Hathaway 2015). 

In our comparative study, we noticed that the magnetic activities such as starspots 
and flaring frequency per rotation found to be increasingly high from F-type to M-type
stars. Although having similar internal structure, this higher level of magnetic 
activities could be explained by their relatively larger thickness of the convection zone 
to the radiation zone and hence gives a stronger dynamo.

%
\vspace{-1.3mm}
\section*{Acknowledgements}
This research has been done under the Indo-Russian DST-RFBR project reference INT/RUS/RFBR/P-167, INT/RUS/RFBR/P-271 (for India) and Grant RFBR Ind\_a 14-02-92694, 17-52-45048  (for Russia). 
This paper also includes data collected by the \textit{Kepler} mission. 
Funding for the \textit{Kepler} mission is provided by the NASA Science Mission directorate.
We acknowledge NASA Exoplanet Archive, All Sky Automated Survey archive, \hipp ~archive, \wasp ~archive, and  different telescope facilities we used to carry out our research. 

%
%
%

\vspace{-0.2cm}
\footnotesize
\beginrefer
\vspace{-0.2cm}

\refer Bailer-Jones C. A. L. et al. 2018, AJ, 156, 58B 

\refer Balona L. A. 2015, MNRAS, 447, 2714

\refer Barnes J. R., Collier Cameron A., Lister T. A., Pointer G. R., Still M. D. 2005, MNRAS, 356, 1501

\refer Benz A. O., G\"{u}del M. 2010, ARA\&A, 48, 241

\refer Collier Cameron A., Donati J.-F. 2002, MNRAS, 329, L23

\refer Csorv{\'a}si R. 2006, Proc. 4th Workshop of Young Researchers in Astronomy \& Astrophysics, 17, 95

\refer Dal H. A., Ta\c{s} G. 2003, IBVS, 5390, 1

\refer David T. J. et al. 2016, Nature, 534, 658

\refer Eibe M. T., Byrne P. B., Jeffries R. D., Gunn A. G. 1999, A\&A, 341, 527

\refer Favata F. et al. 2000, 353, 987


\refer Fluri D. M., Berdyugina S. V. 2004, SoPh, 224, 153

\refer Frasca A., Molenda-Zakowicz J., De Cat P.  et al. 2016, A\&A, 594, 39

\refer Gaia Collaboration, Brown A. G. A et al. 2018, A\&A, 616, A1

\refer Hathway D. H. 2015, LRSP, 2015, 12, 4

\refer Horne J. H., Baliunas S. L. 1986, ApJ, 302, 757

\refer Jeffries R. D., Byrne P. B., Doyle J. G., Anders G. J., James D. J., Lanzafame A. C., 1994, MNRAS, 270, 153

\refer Karmakar S. et al. 2016, MNRAS, 459, 3112

\refer Karmakar S. et al. 2017, ApJ, 840, 102

\refer Karmakar S. et al. 2018, Proc. IAU Symp., 340, 229

\refer Korhonen H., Berdyugina S. V., Tuominen I. 2002, A\&A, 390, 179

\refer Kurucz R. L. 1992, in Barbuy B., Renzini A., Proc. IAU Symp., 149, 225

\refer Lister T. A., Collier Cameron A., Bartus J. 1999, MNRAS, 307, 685

\refer Luo et al. 2016, yCat, 5149, 0

\refer Mann et al. 2016, AJ, 152, 61 

\refer Mann et al. 2017, AJ, 153, 64 

\refer McQuillan et al. 2014, ApJS, 211, 24

\refer Messina S., Guinan E. F. 2003, A\&A, 409, 1017

\refer Pandey J. C., Singh K. P., Drake S. A., Sagar R. 2005, AJ, 130, 1231

\refer Pecaut M. J. et al. 2012, ApJ, 746, 154

\refer Piluso N., Lanza A. F., Pagano I., Lanzafame A. C., Donati J.-F. 2008, MNRAS, 387, 237

\refer Press W. H., Rybicki G. B. 1989, ApJ, 338, 277

\refer Rebull L. M. et al. 2017, ApJ, 839, 92R 

\refer Savanov I. S.  et al. 2016, AcA, 66, 381

\refer Savanov I. S.  et al. 2018, ARep, 62, 532 

\refer Savanov I. S., Strassmeier K. G. 2008, AN, 329, 364

\refer Scargle J. D. 1982, ApJ, 263, 835 

\refer Ta\c{s} G. 2011, AN, 332, 57

\endrefer           

\end{document}